\documentclass[]{pasj01}
\usepackage[switch,mathlines]{lineno}

\Received{}
\Accepted{}
 
\begin{document} 

\title{ 
KIC 7914906: An Eclipsing Heartbeat Star with Tidally Excited Oscillations and Gamma Doradus/Delta Scuti Hybrid Pulsations}

\author{Min-Yu \textsc{Li}\altaffilmark{1,2}}
\altaffiltext{1}{Yunnan Observatories, Chinese Academy of Sciences, Kunming 650216, People's Republic of China}
\altaffiltext{2}{University of Chinese Academy of Sciences, No.1 Yanqihu East Road, Huairou District, Beijing 101408, People's Republic of China}
\email{liminyu@ynao.ac.cn}

\author{Sheng-Bang \textsc{Qian},\altaffilmark{3,4}}
\altaffiltext{3}{Department of Astronomy, School of Physics and Astronomy, Yunnan University, Kunming 650091, People's Republic of China}
\altaffiltext{4}{Key Laboratory of Astroparticle Physics of Yunnan Province, Yunnan University, Kunming 650091, People's Republic of China}

\author{Li-Ying \textsc{Zhu}\altaffilmark{1,2}}

\author{Liang \textsc{Liu}\altaffilmark{1,2}}
\author{Wen-Ping \textsc{Liao}\altaffilmark{1,2}}
\author{Er-Gang \textsc{Zhao}\altaffilmark{1,2}}
\author{Xiang-Dong \textsc{Shi}\altaffilmark{1,2}}
\author{Fu-Xing \textsc{Li}\altaffilmark{1,2}}
\author{Qi-Bin \textsc{Sun}\altaffilmark{3,4}}
 


\maketitle

\begin{abstract}
We present the eclipsing Heartbeat Star KIC 9704906 with tidally excited oscillations (TEOs) and gamma Doradus/delta Scuti hybrid pulsations. The derived parameters show that it has an orbital period of $P$=8.7529108(1) days, a high eccentricity of $e$=0.467(3), and a high inclination of $i$=78$^{\circ}$.81(6). The mass ratio $q$=0.981(5), the relative radii (radius divided by semi-major axis) $r_1$=0.0639(2), and $r_2$=0.0715(4) indicate that the secondary component has a less mass and a larger radius, and may have evolved off the main sequence. The eight derived TEO candidates, $n$ = 3, 4, 5, 6, 7, 12, 40, and 44 harmonics, are consistent with or close to the dominant spherical harmonic $l=2$, $m=0$, or $\pm2$, assuming that the spin and orbital axes are aligned, and the pulsations are adiabatic and standing waves. We also identify eight independent frequency candidates, but one of them, $\mathit{f_{7}}$, is more like a modulation of a quasi-periodic signal and the orbits. According to the g-mode frequencies, we find that the rotation period of one component is 11.52(29) days. Although the masses and radii cannot be further constrained due to the lack of sufficient high-precision spectra, the fascinating phenomena in the Fourier spectra are evident and valuable in this system.
\end{abstract}

\section{Introduction} \label{sec:intro}
Pulsations are a powerful probe of the internal structure of stars using asteroseismology. The gamma Doradus ($\gamma$ Dor) g mode, driven by the convective blocking mechanism \citep{2000ApJ...542L..57G, 2005A&A...435..927D}, and the delta Scuti ($\delta$ Sct) p mode, excited by the $\kappa$ mechanism in the He {\footnotesize II} ionization zone \citep{1995ARA&A..33...75G, 2001A&A...366..178R}, are the two more common categories and they have been studied extensively \citep{2015AJ....149...68B, 2018MNRAS.475..478Q, 2019RAA....19....1Q, 2023A&A...672A.183A}. In addition, several systems exhibiting both $\gamma$ Dor and $\delta$ Sct properties have been reported \citep{2002MNRAS.333..251H, 2010ApJ...713L.192G, 2013A&A...556A..87C, 2014A&A...563A..59M, 2015MNRAS.452.3073B, 2016MNRAS.457.3163X, 2016A&A...592A.116S, 2017ApJ...851...39G, 2017A&A...602A..30H, 2018A&A...610A..17L, 2021A&A...655A..63Y, 2024ApJ...963..155C}.

On the other hand, the tidally excited oscillations (TEOs; \citet{1975A&A....41..329Z,1995ApJ...449..294K,2017MNRAS.472.1538F}) in Heartbeat Star (HBS) systems \citep{2012ApJ...753...86T} are induced by strong tidal effects during periastron passage in eccentric orbits. The main characteristic of TEOs is that they occur at exact integer multiples of the orbital frequency \citep{2017MNRAS.472.1538F}. So far, a number of papers have reported the TEOs in HBS systems \citep{2011ApJS..197....4W, 2012MNRAS.420.3126F, 2012MNRAS.421..983B, 2014MNRAS.440.3036O, 2016MNRAS.463.1199H, 2017ApJ...834...59G, 2019MNRAS.489.4705J, 2021MNRAS.506.4083J, 2020ApJ...896..161G, 2020ApJ...888...95G, 2020ApJ...903..122C, 2021A&A...647A..12K, 2022A&A...659A..47K, 2022ApJ...928..135W, 2022MNRAS.517..437G, 2023AJ....166...42W, 2024ApJ...962...44L}. 

However, pulsating stars with TEOs and $\gamma$ Dor/$\delta$ Sct hybrids are rare and interesting objects. \citet{2013MNRAS.434..925H} reported KIC 4544587 with $\delta$ Sct pulsations and TEOs. They identified 31 modes, including eight TEOs, six anharmonic g-mode $\gamma$ Dor pulsation candidates, and 17 p-mode pulsations. \citet{2019ApJ...885...46G} studied the TEOs, $\delta$ Sct p modes, and the self-driven $\gamma$ Dor g modes in KIC 4142768. They derived the fundamental stellar properties of the HBS, which are in agreement with all the observations.

This paper presents the eclipsing HBS KIC 7914906, which exhibits TEOs and $\gamma$ Dor/$\delta$ Sct oscillations. Section \ref{sec:modeling} describes the binary modeling using the Kepler photometric data and the known atmospheric parameters. Section \ref{sec:pulsation} performs a Fourier analysis of the residual light curve and discusses the pulsations, including the investigation of the TEOs, Gamma Doradus-type g modes, and Delta Scuti-type p and g modes. Section \ref{sec:conclusions} summarizes and concludes our work.


\section{Binary Modeling} \label{sec:modeling}
KIC 7914906 was reported as an HBS by \citet{2016AJ....151...68K} based on the long-term baseline and high-precision Kepler \citep{2010Sci...327..977B} photometric data. They provide the raw and detrended data for $\sim$30 minutes (long) cadence and raw $\sim$1 minute (short) cadence data spanning from May 2009 to May 2013 in the online catalog\footnote{http://keplerEBs.villanova.edu}. We download the photometric data file and use the detrended data \citep{2016AJ....151...68K} of all 18 quarters for our analysis.

\begin{figure}
	\centering
	\includegraphics[width=0.4\textwidth]{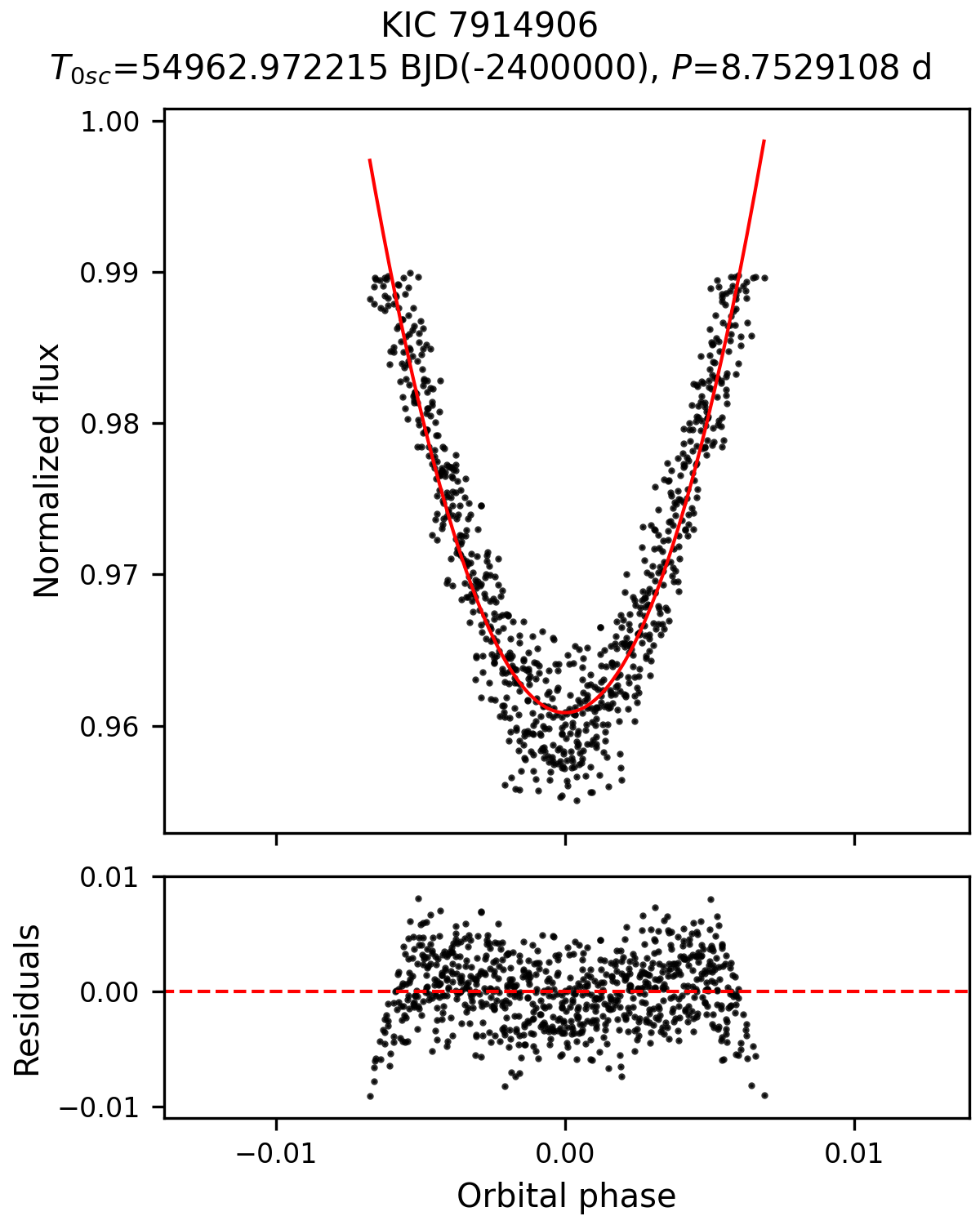}
	\caption{The upper panel shows the eclipse parts of the light curves (black dots) fit to a parabola (red solid line). The lower panel shows the residuals of the fit.
		\label{fig:t0_p}}
\end{figure}

In the first step, we fit the eclipsing parts of the light curves to a parabola, $y=a(\frac{x-T_{0sc}}{P})^2+c$, where $T_{0sc}$ is the time of superior conjunction, and $P$ is the orbital period. As shown in Fig. \ref{fig:t0_p}, the black dots with phase range of (-0.01, 0.01) and the normalized flux less than 0.99 are considered as the eclipsing parts. We then obtain $P$=8.7529108(1) days and $T_{0sc}$=54962.972215(6) (BJD$-$2,400,000), which is very slightly different from that of \citet{2016AJ....151...68K} ($P$=8.7529074, $T_{0sc}$=54962.971405). 

In the second step, we remove the eclipsing parts and fit the remaining parts of the light curves to the corrected \citet{1995ApJ...449..294K} model (K95$^+$ model) following our previous work \citep{2023ApJS..266...28L}. Although the K95$^+$ model is not suitable for eclipsing HBSs, we suggest that the eccentricity obtained from this approach is proper. We obtain an eccentricity $e$=0.518, which can be used for further constraints in the following step.

In the absence of sufficient high-precision spectra, we need to gather useful constraints on this system. \citet{2014MNRAS.437.3473A} have derived the effective temperatures, $T_{\rm eff1}$=7227$\pm$365 K and $T_{\rm eff2}$=7223$\pm$651 K, by fitting the spectral energy distribution. \citet{2014ApJS..211....2H} have obtained the effective temperature $T_{\rm eff}$=7279$^{+231}_{-322}$ K, the mass $M$=1.800$^{+0.273}_{-0.343}$ M$_{\odot}$, and the radius $R$=2.624$^{+0.884}_{-0.778}$ R$_{\odot}$ based on broadband photometry and asteroseismology. The Large Sky Area Multi-Object Fiber Spectroscopic Telescope (LAMOST, also called the Guo Shou Jing Telescope) \citep{1996ApOpt..35.5155W,2015RAA....15.1095L} has provided three low-resolution spectra. The atmospheric parameters are listed in Table \ref{tab:LAMOST}. 

\begin{table}
	\tbl{Atmospheric Parameters from LAMOST.}{%
		\begin{tabular}{lccc}
			\hline
			Parameter & pipe line 1 & pipe line 2 & pipe line 3  \\ 
			\hline
			$T_{\rm eff}$ (K) & 7207(26) & 7382(80) & 7209(16) \\ 
			log g & 4.005(36) & 3.869(130) & 4.034(20) \\ 
			$[$Fe/H$]$ & -0.122(21) & -0.055(86) & -0.087(17) \\
			RV (km/s) & -22.33(342) & -24.52(665) & -17.78(285) \\
			Spec-Type & F0 & F0 & F0 \\
			DATE-OBS & 2013-10-05 & 2014-05-22 & 2015-10-08 \\
			 & T12:28:00 & T19:50:00 & T11:52:00 \\
			Phase & 0.72 & 0.91 & 0.46 \\
			\hline
	\end{tabular}}\label{tab:LAMOST}
	\begin{tabnote}
		The Spec-Type row is the spectral type. The DATE-OBS is the median observation time UTC. The phase is calculated according to the $P$ and $T_{0sc}$ derived in the first step. These results can be found at https://www.lamost.org/dr7/.  \\ 
	\end{tabnote}
\end{table}

In the third step, we apply the binary model code PHysics Of Eclipsing BinariEs (PHOEBE, \citet{2016ApJS..227...29P, 2018maeb.book.....P}) to the light curves using the emcee v3.1.4 Python package \citep{2013PASP..125..306F} based on the Markov Chain Monte Carlo (MCMC) method.  Given the above atmospheric parameters, as well as the similar temperatures from \citet{2014MNRAS.437.3473A}, and assuming that the two stars were born at similar times, we propose that both components are main sequence, with similar surface temperatures and similar masses. We also assume that the hotter component has an effective temperature of 7260 K, the mean of the effective temperatures above, and will fix it in the MCMC run. We further find that the secondary is the hotter component in this scenario, after trying several combinations of superparameters in the fitting procedures. Therefore, we fix the secondary temperature in the MCMC run.

In the first round of fitting, the priors of the parameters are set to uniform distributions. We fix $T_{\rm eff2}$ at 7260 K and set the temperature ratio $tratio$=$T_{\rm eff2}$/$T_{\rm eff1}$ in the range (1.0, 1.1). We set the mass ratio $q$=M${_2}$/M${_1}$ in the range (0.9, 1.1), since we make no assumptions about which of the components is more massive. The eccentricity $e$ is set in the range (0.4, 0.6) according to the result of the second step. The inclination $i$ is set in the range (75$^{\circ}$, 85$^{\circ}$) because of the obvious eclipse in the light curves. The argument of periastron $\omega$ is set in the range (0$^{\circ}$, 180$^{\circ}$). Due to the lack of sufficient radial velocities (RVs), we cannot determine either the absolute mass or the semi-major axis (sma) of the orbit from the light curves alone. We just fix the sma to a unit value and fit the relative radii $r_1$ and $r_2$, where $r_1$ and $r_2$ are the primary and secondary radii divided by the sma, respectively. The ranges of $r_1$ and $r_2$ are set in the range (0.06, 0.08) according to a rough calculation of Kepler's third law based on the mass and radius reported by \citet{2014ApJS..211....2H}. The $T_{0sc}$ is set in the range (54962.97, 54962.98), a small range around the value derived in the first step. Both bolometric albedos, $albedo_1$ and $albedo_2$, are set in the range (0.6, 0.8) according to the effective temperatures. Both gravity darkening coefficients, $gdc_1$ and $gdc_2$, are set in the range (0.45, 1.0) according to the radiative envelope. The limb darkening coefficients are set in interpolated mode. The $P$ is fixed at the value derived in the first step. Finally, we will simultaneously fit twelve parameters, $e$, $i$, $\omega$, $T_{0sc}$, $q$, $tratio$, $r_1$, $r_2$, $albedo_1$, $albedo_2$, $gdc_1$, and $gdc_2$ in the MCMC run.

\begin{figure*}
	\begin{center}
	\includegraphics[width=1.0\textwidth]{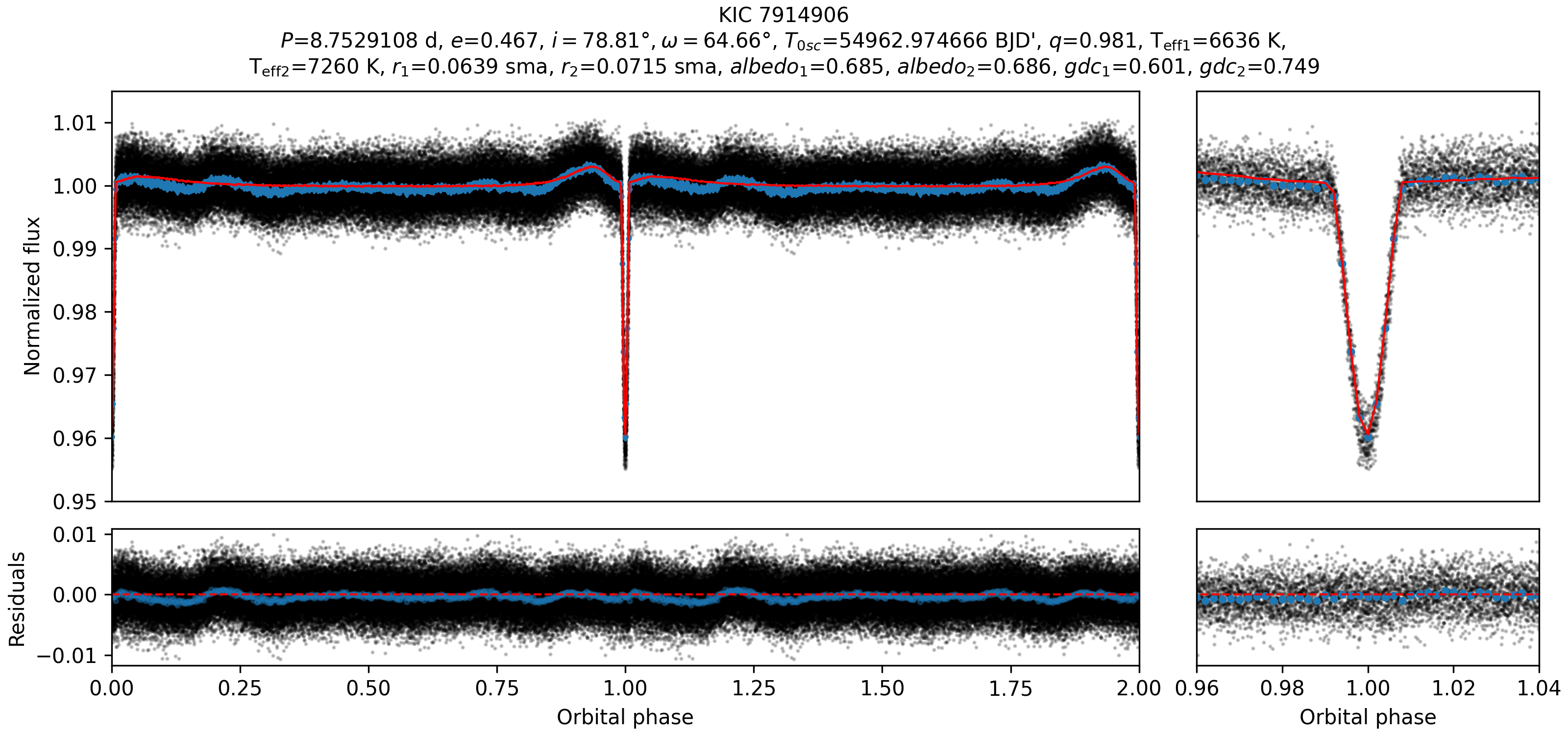}
	\end{center}
	\caption{Fitting results for KIC 7914906. The PHOEBE model (solid red line) fitted to the phase-folded light curve (black dots) in the upper-left panel. The blue dots represent the phase-binned data. The lower-left panel shows the residuals of the fit. The dashed red line indicates a zero point. Note: The plot shows phases 0$-$2 to make the fitting results clear, and phases 1$-$2 are an exact copy of phases 0$-$1. The two right panels show zoomed light curves around eclipse.
		\label{fig:fitrst}}
\end{figure*}

Since the fitting using the PHOEBE model is a very time-consuming procedure, we use 500 bins in phase to reduce the computation. In the MCMC run, we use 32 walkers and 5000 steps for a single chain. Then we obtain the convergence parameters. We then use the same walkers and steps in a single chain to derive the parameter uncertainties. The priors are set to be a normal distribution centered on the value from the previous step. Fig. \ref{fig:fitrst} shows the fitting results, and Fig. \ref{fig:corner} shows the corner plot of the MCMC fitting procedure. The parameters of the binary model are listed in Table \ref{tab:par}.

\begin{table}
	\tbl{Binary Model Parameters.}{%
		\begin{tabular}{lr}
			\hline
			Parameter & Value\\ 
			\hline
			Orbital Period, $P$ (days) & 8.7529108(1) \\
			Time of Superior Conjunction, $T_{0sc}$ (BJD')& 54962.974666(50) \\
			Time of Periastron, $T_{0p}$ (BJD') & 54962.772581(50) $^b$ \\
			eccentricity, $e$ & 0.467(3) \\
			inclination, $i$ (degree) & 78.81(6) \\
			Argument of periastron, $\omega$ (degree) & 64.66(7) \\
			Mass ratio (M${_2}$/M${_1}$), $q$ & 0.981(5) \\
			Temperature ratio ($T_{\rm eff2}$/$T_{\rm eff1}$), tratio & 1.094(3)\\
			$T_{\rm eff1}$ (K) & 6636(20) \\ 
			$T_{\rm eff2}$ (K) & 7260 $^a$ \\
			$r_1$ (sma) & 0.0639(2) \\
			$r_2$ (sma) & 0.0715(4) \\
			$albedo_1$ & 0.685(4) \\
			$albedo_2$ &0.686(5) \\
			$gdc_1$ & 0.601(3) \\
			$gdc_2$ & 0.749(2) \\
			\hline
	\end{tabular}}\label{tab:par}
	\begin{tabnote}
		The units of $T_{0sc}$ and $T_{0p}$ are BJD'=BJD$-$2,400,000. \\
		$^a$ fixed. \\
		$^b$ derived by the PHOEBE function.
	\end{tabnote}
\end{table}

\section{Pulsation Analysis} \label{sec:pulsation}
\begin{figure*}
	\begin{center}
	\includegraphics[width=1.0\textwidth]{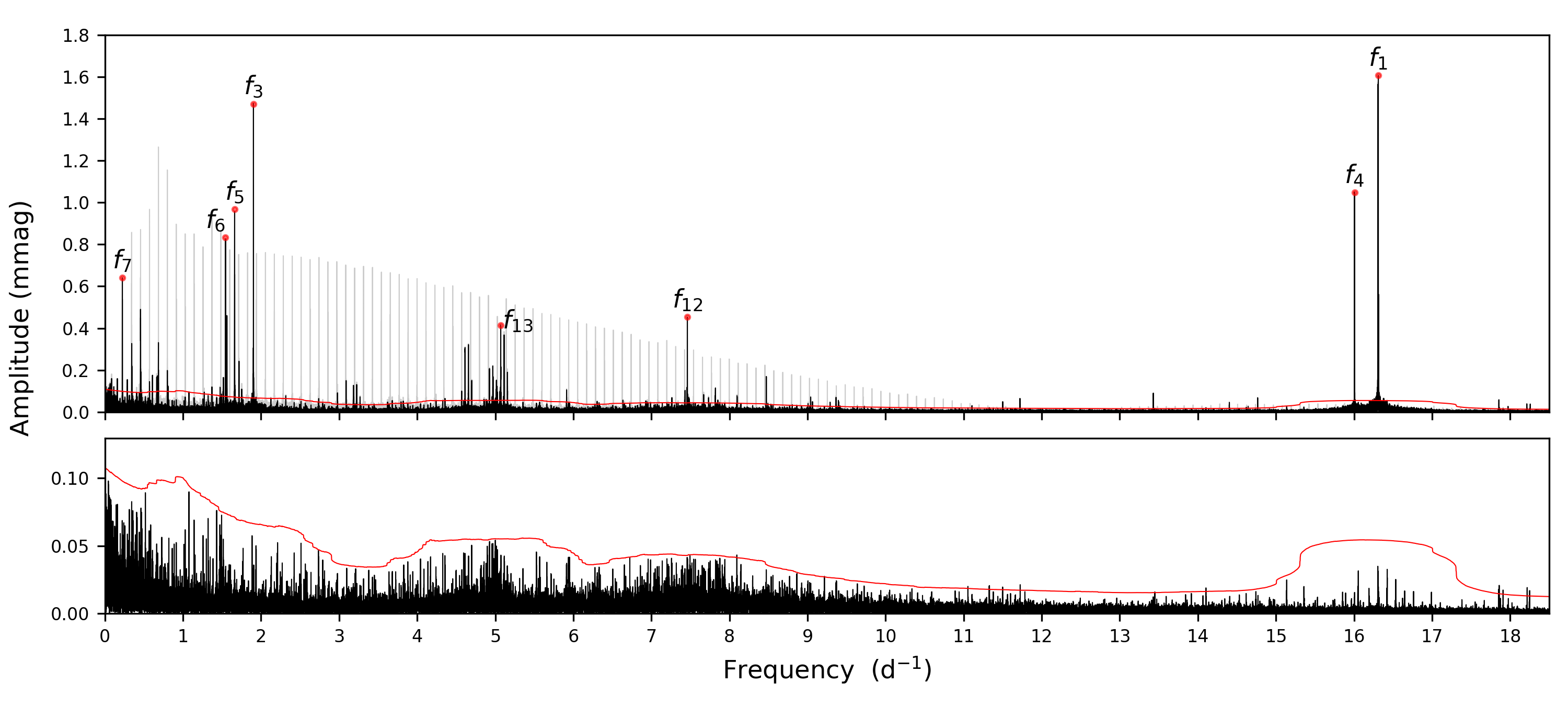}
	\end{center}
	\caption{Upper panel: the Fourier spectrum of the residuals of the fit (black dots in the lower-left panel of Fig. \ref{fig:fitrst}). For comparison, the light gray curve is the Fourier spectrum of the original data (black dots in the upper-left panel of Fig. \ref{fig:fitrst}). The independent frequencies in Table \ref{tab:indep} are also labeled. Lower panel: the residual spectrum after removing the 209 detected frequencies. The red lines in both panels represent the amplitudes at $S/N$ = 4.0 as a function of frequency.
		\label{fig:spectra}}
\end{figure*}

We perform the Fourier spectral analysis on the residuals of the light curve data, the black dots in the lower-left panel of Fig. \ref{fig:fitrst}, using the FNPEAKS \footnote{http://helas.astro.uni.wroc.pl/deliverables.php?active=fnpeaks} code-based analytic procedures. The flux variation is formulated as a sinusoidal equation:
\begin{equation}\label{eqs:ft}
	\Sigma A_i {\rm sin} [2\pi(f_i \cdot t+\phi_i)]
\end{equation}
where $A_i$, $f_i$, and $\phi_i$ are the amplitude, frequency, and phase, respectively. The mean noise level of each frequency, $N_i$, is derived as the mean amplitude in the frequency range $\pm$1 d$^{-1}$. The signal-to-noise ratio ($S/N$) of the frequencies is $A_i/N_i$. We iteratively extract the peak frequencies until the $S/N$ is no greater than 4.0 using Eq. (\ref{eqs:ft}) \footnote{The fnpeaks code does not support phase derivation. However, it already derives the real and imaginary parts of the complex value of the Fourier transform results, and the phases can be calculated with minor modifications. The modified code can be downloaded from https://github.com/MinyuLi/fnpeaks$\_$phase.}. Uncertainties of frequencies, amplitudes and phases are estimated following \citet{2008A&A...481..571K} (their Eqs. (1)$\sim$(4)). In addition, the time of periastron, $T_{0p}$, should be subtracted from the time of each data point in the photometric light curves, before Fourier analysis, where $T_{0p}$=54962.772581 can be derived by the PHOEBE function according to the parameters obtained in Section \ref{sec:modeling}. The purpose of adjusting the time of data points is to derive the correct phases of TEOs for the mode identification in Section \ref{sec:teos} \citep{2024MNRAS.530..586L}. Finally, a total of 209 frequencies are detected. Table \ref{tab:indep} shows the eight independent frequency candidates; Table \ref{tab:teos} shows the eight TEOs; Table \ref{tab:freq} shows the other 193 combination frequencies or side lobes. When two frequencies are closer than the frequency resolution 1.5/$T_{span}$$\simeq$0.00103 d$^{-1}$, where $T_{span}$ is the time span of about 1462 days of the Kepler data points, the frequency with the lower amplitude is identified as the possible side lobe. The combination frequencies in Table \ref{tab:freq} also satisfy the requirement that the difference between the observed frequency and the value calculated by the equation in the Remark column is less than 1.5/$T_{span}$ (Only one exception: The $\simeq$ in the Remark column of $\mathit{f_{18}}$ is because the difference is about 0.005, which is greater than 1.5/$T_{span}$.). Fig. \ref{fig:spectra} shows the Fourier spectra. The Fourier spectrum of the residuals of the fit (black dots in the lower-left panel of Fig. \ref{fig:fitrst}) is shown in the upper panel. For comparison, the light gray curve is the Fourier spectrum of the original data (black dots in the upper-left panel of Fig. \ref{fig:fitrst}). The independent frequencies are also labeled. The residual spectrum after removing the 209 detected frequencies is shown in the lower panel.

\begin{table}
	\tbl{Independent Frequency Candidates.}{%
		\begin{tabular}{lcccc}
			\hline
			ID & Freq.(d$^{-1}$) & Ampl.(mmag) & S/N & Remark  \\ 
			\hline
			$\mathit{f_{1}}$ & 16.3109249(20) & 1.6070(84) & 118.50 &  $\delta$ Sct   \\ 
			$\mathit{f_{3}}$ & 1.9028370(26) & 1.472(10) & 88.65 &   $\gamma$ Dor  \\ 
			$\mathit{f_{4}}$ & 16.0065308(30) & 1.0485(84) & 77.24 &   $\delta$ Sct \\ 
			$\mathit{f_{5}}$ & 1.6621303(43) & 0.968(11) & 54.33 &   $\gamma$ Dor \\ 
			$\mathit{f_{6}}$ & 1.5449563(52) & 0.833(11) & 45.07 &   $\gamma$ Dor \\ 
			$\mathit{f_{7}}$ & 0.2231611(89) & 0.641(15) & 26.20 &  ? \\ 
			$\mathit{f_{12}}$ & 7.4603409(55) & 0.4526(66) & 42.16 & $\delta$ Sct \\ 
			$\mathit{f_{13}}$ & 5.0701402(78) & 0.4126(85) & 29.94 & g-mode \\ 
			\hline
	\end{tabular}}\label{tab:indep}
	\begin{tabnote}
		The question mark in the Remark column of $\mathit{f_{7}}$ indicates that it may not be an independent frequency, see text in Section \ref{sec:gamma-dor}. 
	\end{tabnote}
\end{table}

\begin{table}
	\tbl{Tidally Excited Oscillations.}{%
		\begin{tabular}{lcccccc}
			\hline
			ID & Freq. & Ampl. & Phase & S/N & $f$/$f_{\rm orb}$ & n  \\ 
			 & (d$^{-1}$)& (mmag) &   &   &   & \\
			\hline
			$\mathit{f_{9}}$ & 0.456217(11) &  0.493(14) & 0.164(9) & 21.39 & 3.993 & 4 \\ 
			$\mathit{f_{15}}$ & 0.685465(17) & 0.332(15) & 0.274(15) & 13.46 & 6.000 & 6 \\ 
			$\mathit{f_{16}}$ & 0.343769(17) & 0.331(15) & 0.760(14) & 14.04 & 3.009 & 3 \\ 
			$\mathit{f_{32}}$ & 0.799714(28) & 0.201(15) & 0.219(24) & 8.26 & 7.000 & 7 \\ 
			$\mathit{f_{49}}$ & 0.571215(38) & 0.147(15) & 0.478(32) & 6.08 & 5.000 & 5 \\ 
			$\mathit{f_{69}}$ & 1.370997(39) & 0.121(13) & 0.005(33) & 5.95 & 12.000 & 12 \\ 
			$\mathit{f_{84}}$ & 4.570127(31) & 0.1038(85) & 0.843(26) & 7.58 & 40.002 & 40 \\ 	
			$\mathit{f_{87}}$ & 5.026616(31) & 0.1024(85) & 0.171(27) & 7.42 & 43.998 & 44 \\
			\hline
	\end{tabular}}\label{tab:teos}
	\begin{tabnote}
		Column $n$ is the harmonic number of the TEO.
	\end{tabnote}
\end{table}

\subsection{Tidally Excited Oscillations} \label{sec:teos}

We use the analytic procedure reported in \citet{2024ApJ...962...44L} to detect the TEOs. As shown in the upper panel of Fig. \ref{fig:spectra}, we apply Fourier transforms to the residuals (black curve) as well as to the original light curve (light gray curve). We consider the harmonic $n$ to be a TEO if it occurs in both spectra. We also visually inspect all fits and exclude some frequencies with $S/N<$ 5.0 that may have been injected by imperfect model subtraction. Finally, the TEO candidates are the $n$ = 4, 6, 3, 7, 5, 12, 40, and 44 harmonics.  The harmonic TEO candidates are listed in Table \ref{tab:teos}. 

\begin{figure}
	\begin{center}
		\includegraphics[width=0.5\textwidth]{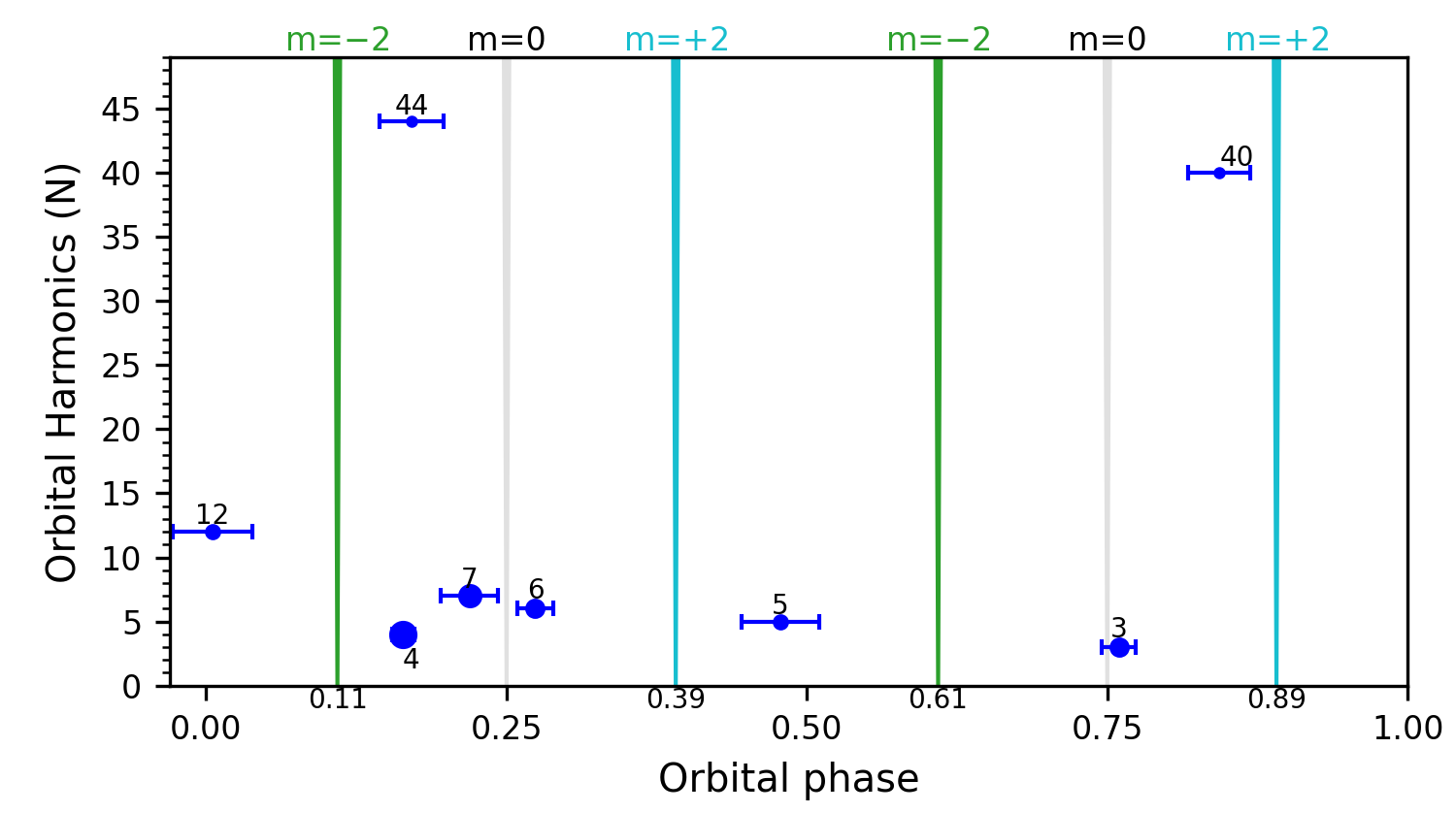} 
	\end{center}
	\caption{The pulsation phases of the TEOs. The gray, light blue, and green strips indicate the $m=0,+2$, and $-2$ modes, respectively. The phases of the $m=+2,-2$ modes are shown near the strips. The blue circle represents a TEO with its harmonic number $n$; the size corresponds to its amplitude; the error bar corresponds to the uncertainty of its phase.}\label{fig:PM}
\end{figure}

We then identify the pulsation phases and mode of these TEOs following \citet{2024MNRAS.530..586L}. The pulsation phases of the TEOs for dominant modes of spherical harmonic degree $l=2$ can be expressed by the equation \citep{2014MNRAS.440.3036O, 2020ApJ...888...95G} :
\begin{equation} \label{equation:phi}
	\phi_{_{l=2,m}}=0.25+m\phi_{_{0}},
\end{equation}
where azimuthal order $m = 0, \pm2$, $\phi_{_{0}}=0.25-\omega/360^{\circ}$ and $\omega$ is the argument of periastron. This equation holds under the following assumptions, including that the spin and orbital axes are aligned and that the pulsations are adiabatic and standing waves. As shown in Fig. \ref{fig:PM}, the gray, light blue, and green strips represent the theoretical pulsation phases $\phi_{_{m}}$ for $m=0,+2$, and $-2$ mode, respectively. The $n$ = 3, 6, and 7 harmonics are consistent with the $m=0$ mode. The $n$ = 4, 5, 12, 40, and 44 harmonics are close to the $|m|=2$ mode.

\subsection{Gamma Doradus-type g Modes} \label{sec:gamma-dor}
\begin{figure*}
	\begin{center}
		\includegraphics[width=1.0\textwidth]{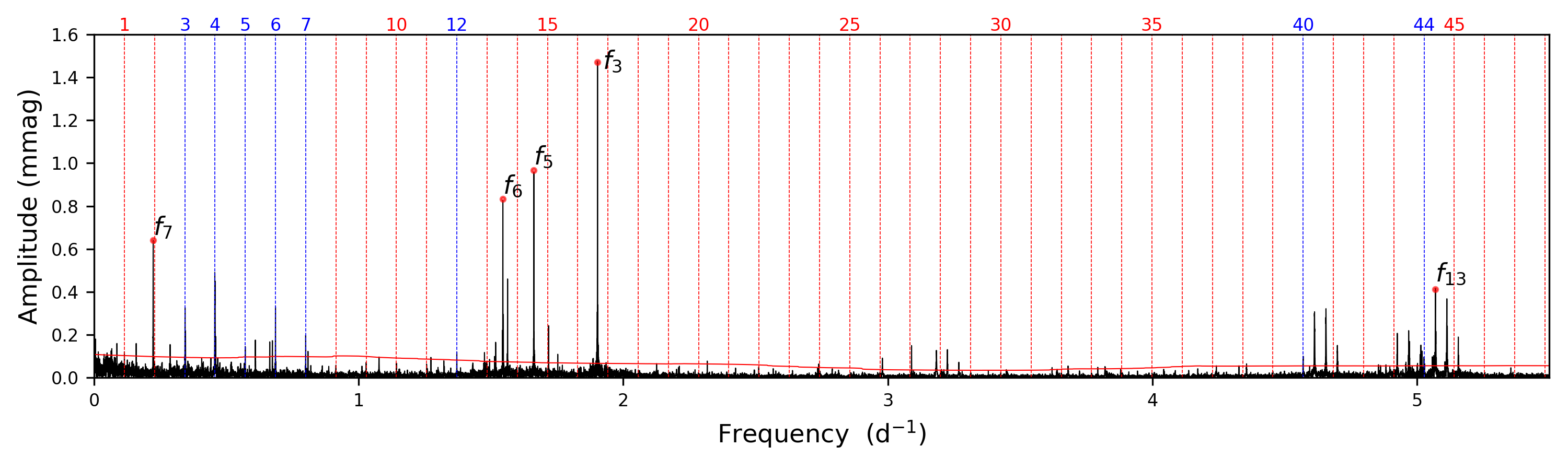}
	\end{center}
	\caption{The Fourier spectrum in the range from 0 to 5.5 d$^{-1}$. The red and blue vertical dashed lines represent the orbital harmonics $n$; the blue ones indicate that they are TEOs. The solid red line shows the amplitudes at $S/N$ = 4.0 as a function of frequency. The independent frequency candidates less than 6 d$^{-1}$ in Table \ref{tab:indep} are also labeled.
		\label{fig:teos}}
\end{figure*}

$\gamma$ Dor stars are characterized by low-frequency g-mode pulsations with masses ranging from 1.3 to 2.0 M$_{\odot}$ and periods of about 0.3 to 3 days \citep{2019RAA....19....1Q, 2019ApJ...885...46G}. As can be seen in Fig. \ref{fig:teos} and Table \ref{tab:indep}, $\mathit{f_{3}}$, $\mathit{f_{5}}$, $\mathit{f_{6}}$, and $\mathit{f_{7}}$ are the self-excited $\gamma$ Dor frequencies candidates. Since $\mathit{f_{11}}$ is too close to $\mathit{f_{6}}$, we suggest that it is not an independent frequency.

\begin{figure}
	\begin{center}
		\includegraphics[width=0.5\textwidth]{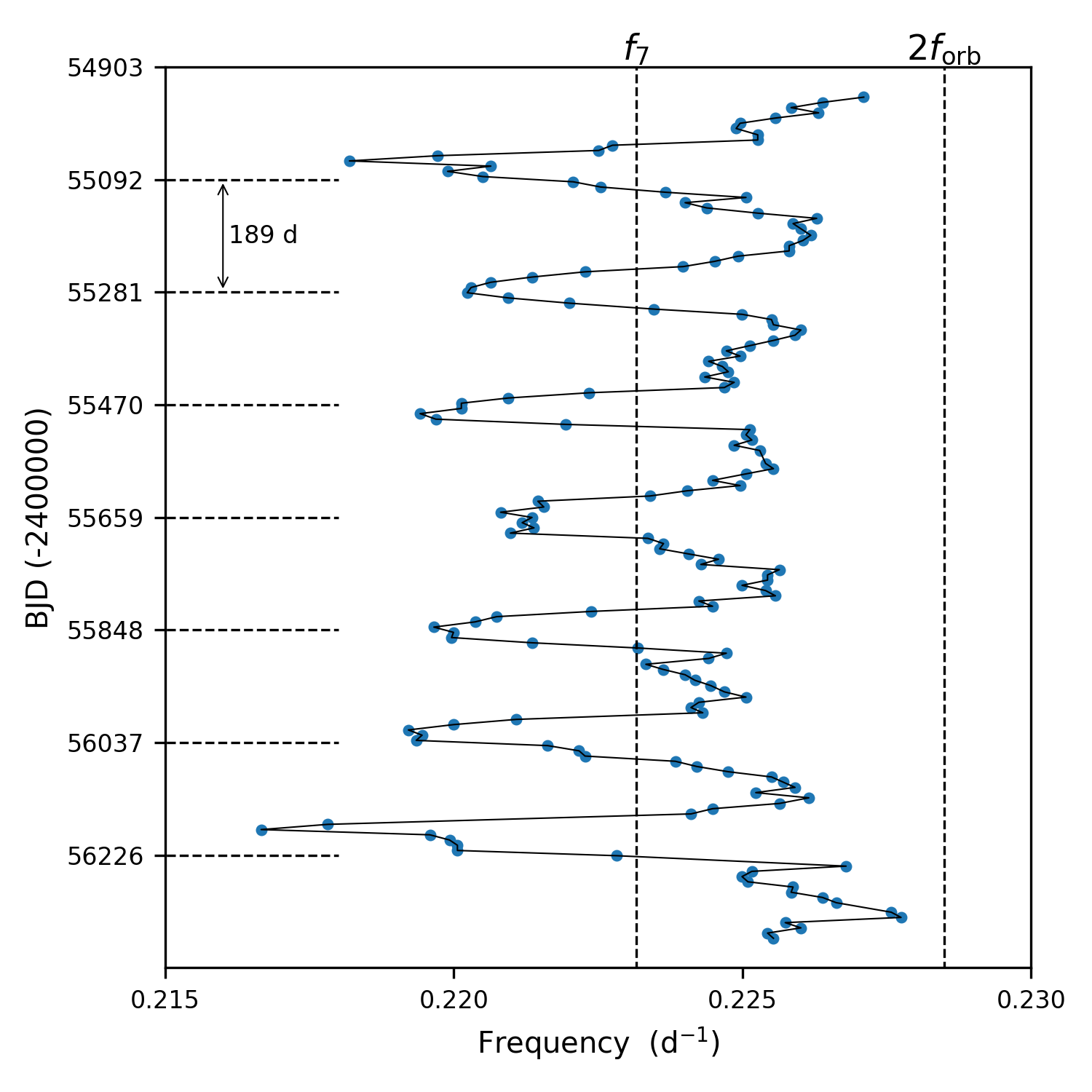} 
	\end{center}
	\caption{Change in peak value of $\mathit{f_{7}}$ over time. The blue dots represent the frequency peak for each data segment. The two vertical dashed lines represent the  $\mathit{f_{7}}$ value in Table \ref{tab:indep} and the 2$f_{\rm orb}$, respectively. The horizontal dashed lines represent the quasi-periodic interval of 189 days.}\label{fig:f7}
\end{figure}

The $\mathit{f_{7}}$$\simeq$1.95$f_{\rm orb}$ is not an exact integer multiple of the orbital frequency. However, since it is easily related to the orbit, we examine whether the frequency changes during all quarters. We slice the photometric data into segments, each starting at an interval of one orbital period and spanning 87 days (about ten orbital periods). We then extract the frequency peak around (0.21 d$^{-1}$, 0.23 d$^{-1}$) for each data segment by Fourier transform. As shown in Fig. \ref{fig:f7}, the frequencies are changing. In addition, there is a quasi-periodicity in their changes. We also perform a Fourier transform on these peak frequencies and find that the quasi-periodic interval is about 189 days. We therefore suggest that $\mathit{f_{7}}$ is a modulation of some quasi-periodic signal and the orbits, rather than an independent frequency.

Furthermore, no changes were found in $\mathit{f_{3}}$, $\mathit{f_{5}}$, and $\mathit{f_{6}}$ after the same examination. Therefore, they are independent frequencies.

\subsection{Delta Scuti-type p and g Modes} \label{sec:delta-sct}
The $\delta$ Sct stars lie in the classical Cepheid instability strip with masses from 1.5 to 3 M$_{\odot}$ and spectral types from A2 to F5. They exhibit both radial and non-radial pulsations with frequencies ranging from 3 to 80 d$^{-1}$ and can oscillate in both p and g modes \citep{2018MNRAS.475..478Q}. The $\mathit{f_{1}}$ and $\mathit{f_{4}}$ in Fig. \ref{fig:spectra} are typical $\delta$ Sct p modes. Since $\mathit{f_{2}}$ and $\mathit{f_{8}}$ are too close to $\mathit{f_{1}}$, we suggest that they are not independent frequencies. In addition, $\mathit{f_{12}}$ and the frequencies around 5 d$^{-1}$ (Figs. \ref{fig:spectra} and \ref{fig:teos}) are g-mode pulsations.

\begin{figure}
	\begin{center}
		\includegraphics[width=0.5\textwidth]{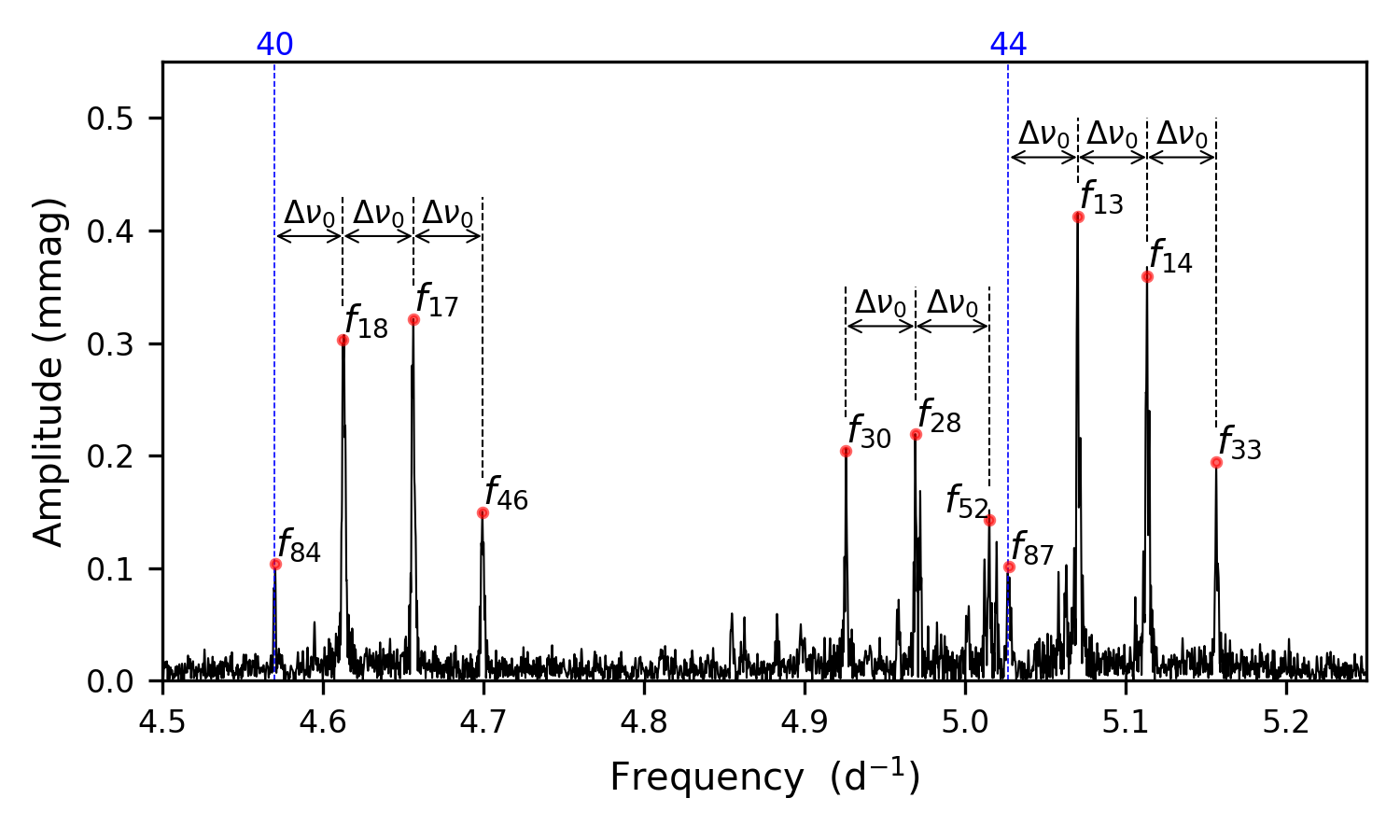} 
	\end{center}
	\caption{The details of the frequency range of 4.5 $\sim$ 5.25 d$^{-1}$.}\label{fig:freq_group}
\end{figure}

However, there are three groups of g-mode frequencies in the frequency range of around 5 d$^{-1}$ that are very noteworthy. Fig. \ref{fig:freq_group} shows the details of these three frequency groups: ($\mathit{f_{84}}$, $\mathit{f_{18}}$, $\mathit{f_{17}}$, $\mathit{f_{46}}$), ($\mathit{f_{30}}$, $\mathit{f_{28}}$, $\mathit{f_{52}}$), and ($\mathit{f_{87}}$, $\mathit{f_{13}}$, $\mathit{f_{14}}$, $\mathit{f_{33}}$). The space between two adjacent frequencies is $\Delta \nu_0$=0.0434(11). Among these frequencies, $\mathit{f_{84}}$ and $\mathit{f_{87}}$ are harmonic TEOs; $\mathit{f_{13}}$ is independent frequency. Given their equal frequency spacing, they may be produced by the rotational splitting effect. Furthermore, these groups of frequencies are likely to be modulations of rotation and orbits. In particular, the splitting of high-order g modes of degree 1 is only half the rotation rate \citep{2010aste.book.....A}. Therefore, we propose that the rotation period of the one component is $P_{rot}$=(2$\Delta \nu_0$)$^{-1}$=11.52(29) d, but it is not yet clear to which component this rotation period belongs. Given the high eccentricity of this system, we expect that at least one component is not tidally locked. That the rotation period is different from the orbital period is not surprising. 

\section{Summary and Conclusions}\label{sec:conclusions}
In this work, we study the eclipsing HBS KIC 7914906. We apply the PHOEBE binary model code to the light curves using the MCMC method. The derived orbital period $P$=8.7529108(1) days, eccentricity $e$=0.467(3), inclination $i$=78$^{\circ}$.81(6), mass ratio $q$=0.981(5), $T_{\rm eff1}$=6636(20) K, $T_{\rm eff2}$=7260 K, the relative radii $r_1$=0.0639(2) sma and $r_2$=0.0715(4) sma. This indicates that the secondary is the less massive and hotter component with a larger radius, and may have evolved off the main sequence.

The Fourier spectral analysis shows that KIC 7914906 exhibits the TEOs, $\gamma$ Dor and $\delta$ Sct hybrid pulsations. We extract 209 frequencies with $S/N$ $>$ 4.0, including eight independent frequency candidates, eight TEO candidates, and 193 combination frequencies or side lobes. According to the pulsation phases and mode identification, the $n$ = 3, 6, and 7 harmonics are spherical harmonic degree $l=2$ and azimuthal order $m = 0$ mode; the $n$ = 4, 5, 12, 40, and 44 harmonics are $l=2$ and $|m|=2$ mode. We also find the rotation period of one component, which is 11.52(29) days. The $\mathit{f_{7}}$$\simeq$1.95$f_{\rm orb}$ is also interesting in this system. We find that it is a modulation of a quasi-periodic signal with about 189 days and the orbits.

Since the absence of sufficient high-precision spectra, the absolute masses and radii cannot be further constrained. Nor can the pulsation modes of the independent frequencies be further identified. However, the fascinating phenomena in the Fourier spectra are evident and valuable. We leave high-precision spectroscopic observations for future work, when the power of asteroseismology can reveal more of the internal structure of this system.



\begin{ack}
This work is partly supported by the National Key R$\&$D Program of China (Grant No. 2022YFE0127300), the National Natural Science Foundation of China (Nos. 11933008 and 12103084), the Basic Research Project of Yunnan Province (Grant Nos. 202201AT070092 and 202301AT070352), the Science Foundation of Yunnan Province (No. 202401AS070046), and the Yunnan Revitalization Talent Support Program. We are grateful to \citet{2016AJ....151...68K} for their excellent work and for sharing the results of their work. Funding for the Kepler mission is provided by the NASA Science Mission directorate. The Guoshoujing Telescope (the Large Sky Area Multi-Object Fiber Spectroscopic Telescope LAMOST) is a National Major Scientific Project built by the Chinese Academy of Sciences. We thank the Kepler and LAMOST teams for their support and hard work. We thank the anonymous referee for constructive comments that have been a source of improvement for this manuscript.
\end{ack}

\appendix
\setcounter{figure}{0}
\setcounter{table}{0}
\renewcommand{\thefigure}{A\arabic{figure}}
\renewcommand{\thetable}{A\arabic{table}}
\begin{figure*}
	\centering
	\includegraphics[width=1.0\textwidth]{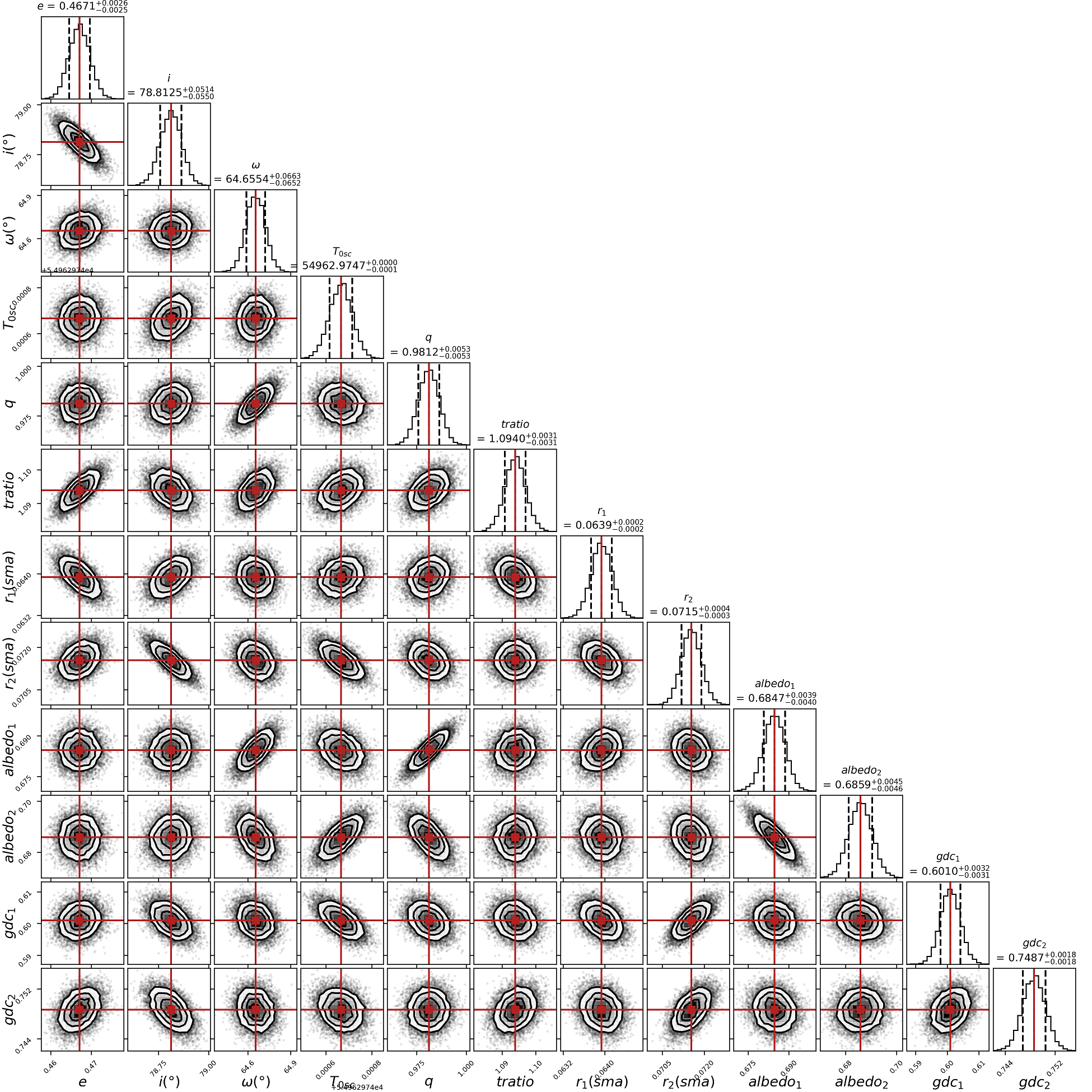}
	\caption{Corner plot of the MCMC fit procedure for KIC 7914906. Red vertical lines indicate the median values of the presented histograms for each parameter. Black vertical dashed lines show 1 $\sigma$ uncertainties. 
		\label{fig:corner}}
\end{figure*}

\section*{}
The corner plot of the MCMC fitting procedure for KIC 7914906 is shown in Fig. \ref{fig:corner}. The other 193 combination frequencies or side lobes are listed in Table \ref{tab:freq}.
\begin{longtable}{lcccclcccc}
	\caption{Possible Combination or Side Lobe Frequencies.}\label{tab:freq}
	\hline              
	ID& Freq.(d$^{-1}$) & Ampl.(mmag) & S/N & Remark &ID& Freq.(d$^{-1}$) & Ampl.(mmag) & S/N & Remark \\ 
	\endfirsthead
	\hline
	ID& Freq.(d$^{-1}$) & Ampl.(mmag) & S/N & Remark &ID& Freq.(d$^{-1}$) & Ampl.(mmag) & S/N & Remark \\
	\endhead
	\hline
	\endfoot
	\hline
	\endlastfoot
	\hline
$\mathit{f_{2}}$ & 16.3058925(21) & 1.5206(84) & 112.11 & $\mathit{f_{1}}$$-$$\mathit{f_{36}}$ & $\mathit{f_{8}}$ & 16.3148352(60) & 0.5288(84) & 39.00 & $\mathit{f_{1}}$$+$2$\mathit{f_{63}}$ \\ 
$\mathit{f_{10}}$ & 0.457169(11) & 0.485(14) & 21.07 & side lobe & $\mathit{f_{11}}$ & 1.5629099(94) & 0.456(11) & 24.81 & 2$\mathit{f_{11}}$$-$7$\mathit{f_{7}}$ \\ 
$\mathit{f_{14}}$ & 5.1133239(90) & 0.3596(85) & 26.08 & 7$\mathit{f_{49}}$$+$5$\mathit{f_{7}}$ & $\mathit{f_{17}}$ & 4.6562568(100) & 0.3208(85) & 23.41 & $\mathit{f_{14}}-4f_{\rm orb}$ \\ 
$\mathit{f_{18}}$ & 4.612393(11) & 0.3029(85) & 22.03 & $\simeq$$\mathit{f_{13}}-4f_{\rm orb}$ & $\mathit{f_{19}}$ & 4.613107(11) & 0.3020(85) & 21.97 & side lobe \\ 
$\mathit{f_{20}}$ & 5.069528(11) & 0.2949(85) & 21.40 & side lobe & $\mathit{f_{21}}$ & 4.655475(11) & 0.2808(85) & 20.49 & side lobe \\ 
$\mathit{f_{22}}$ & 5.112542(13) & 0.2556(85) & 18.53 & side lobe & $\mathit{f_{23}}$ & 5.071806(13) & 0.2424(85) & 17.59 & 5$\mathit{f_{11}}$$-$4$\mathit{f_{15}}$ \\ 
$\mathit{f_{24}}$ & 1.717997(17) & 0.241(11) & 14.00 & 2$\mathit{f_{6}}$$-$2$\mathit{f_{15}}$ & $\mathit{f_{25}}$ & 5.114820(14) & 0.2345(85) & 17.00 & 5$\mathit{f_{32}}$$+$5$\mathit{f_{7}}$ \\ 
$\mathit{f_{26}}$ & 0.345402(25) & 0.223(15) & 9.45 & 9$\mathit{f_{7}}$$-$$\mathit{f_{5}}$ & $\mathit{f_{27}}$ & 4.613821(14) & 0.2216(85) & 16.12 & 6$\mathit{f_{6}}$$-$$\mathit{f_{17}}$ \\ 
$\mathit{f_{28}}$ & 4.968914(15) & 0.2192(85) & 15.90 & 4$\mathit{f_{24}}$$-$$\mathit{f_{3}}$ & $\mathit{f_{29}}$ & 0.222345(27) & 0.215(15) & 8.77 & side lobe \\ 
$\mathit{f_{30}}$ & 4.925798(16) & 0.2045(85) & 14.94 & 4$\mathit{f_{32}}$$+$5$\mathit{f_{26}}$ & $\mathit{f_{31}}$ & 1.546690(21) & 0.203(11) & 10.97 & 9$\mathit{f_{26}}$$-$$\mathit{f_{11}}$ \\ 
$\mathit{f_{33}}$ & 5.156406(17) & 0.1946(85) & 14.10 & 2$\mathit{f_{14}}$$-$$\mathit{f_{13}}$ & $\mathit{f_{34}}$ & 0.223807(30) & 0.190(15) & 7.77 & side lobe \\ 
$\mathit{f_{35}}$ & 0.457883(29) & 0.188(14) & 8.15 & $\mathit{f_{13}}$$-$$\mathit{f_{18}}$ & $\mathit{f_{36}}$ & 0.005134(35) & 0.180(17) & 6.69 & $\mathit{f_{1}}$$-$$\mathit{f_{2}}$ \\ 
$\mathit{f_{37}}$ & 0.609400(31) & 0.179(15) & 7.43 & 2$\mathit{f_{1}}$$-$2$\mathit{f_{4}}$ & $\mathit{f_{38}}$ & 1.517720(25) & 0.173(12) & 9.25 & 6$\mathit{f_{9}}$$-$2$\mathit{f_{37}}$ \\ 
$\mathit{f_{39}}$ & 0.673768(33) & 0.172(15) & 6.97 & 3$\mathit{f_{7}}$$+$$\mathit{f_{36}}$ & $\mathit{f_{40}}$ & 8.472166(12) & 0.1690(56) & 18.71 & 2$\mathit{f_{32}}$$+$4$\mathit{f_{24}}$ \\ 
$\mathit{f_{41}}$ & 4.656835(19) & 0.1669(85) & 12.18 & side lobe & $\mathit{f_{42}}$ & 0.663941(35) & 0.163(15) & 6.62 & 3$\mathit{f_{7}}$$-$$\mathit{f_{36}}$ \\ 
$\mathit{f_{43}}$ & 0.287223(35) & 0.161(15) & 6.70 & 5$\mathit{f_{3}}$$-$2$\mathit{f_{27}}$ & $\mathit{f_{44}}$ & 0.086061(38) & 0.160(16) & 6.15 & 2$\mathit{f_{14}}$$-$2$\mathit{f_{13}}$ \\ 
$\mathit{f_{45}}$ & 7.460953(16) & 0.1576(66) & 14.68 & side lobe & $\mathit{f_{46}}$ & 4.699270(21) & 0.1499(85) & 10.96 & 3$\mathit{f_{5}}$$-$$\mathit{f_{43}}$ \\ 
$\mathit{f_{47}}$ & 3.089845(14) & 0.1491(55) & 16.68 & 2$\mathit{f_{15}}$$+$$\mathit{f_{24}}$ & $\mathit{f_{48}}$ & 4.969492(22) & 0.1470(85) & 10.65 & side lobe \\ 
$\mathit{f_{50}}$ & 4.611747(22) & 0.1453(85) & 10.57 & side lobe & $\mathit{f_{51}}$ & 0.159032(41) & 0.144(16) & 5.70 & 2$\mathit{f_{7}}$$-$$\mathit{f_{43}}$ \\ 
$\mathit{f_{52}}$ & 5.015022(23) & 0.1428(85) & 10.36 & 8$\mathit{f_{49}}$$+$2$\mathit{f_{7}}$ & $\mathit{f_{53}}$ & 5.071024(23) & 0.1402(85) & 10.17 & side lobe \\ 
$\mathit{f_{54}}$ & 0.066850(44) & 0.140(16) & 5.36 & 8$\mathit{f_{7}}$$-$$\mathit{f_{24}}$ & $\mathit{f_{55}}$ & 7.458709(18) & 0.1393(66) & 12.98 & 5$\mathit{f_{3}}$$-$3$\mathit{f_{15}}$ \\ 
$\mathit{f_{56}}$ & 0.459311(39) & 0.137(14) & 5.95 & 3$\mathit{f_{16}}$$-$$\mathit{f_{49}}$ & $\mathit{f_{57}}$ & 4.972008(24) & 0.1368(85) & 9.91 & 5$\mathit{f_{15}}$$+$$\mathit{f_{6}}$ \\ 
$\mathit{f_{58}}$ & 3.225244(15) & 0.1328(54) & 15.35 & $\mathit{f_{5}}$$+$$\mathit{f_{11}}$ & $\mathit{f_{59}}$ & 4.698556(24) & 0.1309(85) & 9.57 & side lobe \\ 
$\mathit{f_{60}}$ & 3.183659(16) & 0.1264(54) & 14.43 & 6$\mathit{f_{9}}$$+$2$\mathit{f_{7}}$ & $\mathit{f_{61}}$ & 1.475522(35) & 0.125(12) & 6.59 & 7$\mathit{f_{7}}$$-$$\mathit{f_{44}}$ \\ 
$\mathit{f_{62}}$ & 4.699951(26) & 0.1236(85) & 9.04 & side lobe & $\mathit{f_{63}}$ & 0.002244(51) & 0.123(17) & 4.57 & $\mathit{f_{56}}-4f_{\rm orb}$ \\ 
$\mathit{f_{64}}$ & 0.037539(51) & 0.123(16) & 4.62 & 2$\mathit{f_{32}}$$-$$\mathit{f_{11}}$ & $\mathit{f_{65}}$ & 0.808419(46) & 0.122(15) & 5.03 & 3$\mathit{f_{16}}$$-$$\mathit{f_{7}}$ \\ 
$\mathit{f_{66}}$ & 5.019680(26) & 0.1220(85) & 8.85 & 8$\mathit{f_{9}}$$+$2$\mathit{f_{15}}$ & $\mathit{f_{67}}$ & 0.048930(50) & 0.122(16) & 4.62 & 3$\mathit{f_{11}}$$-$3$\mathit{f_{31}}$ \\ 
$\mathit{f_{68}}$ & 4.926444(26) & 0.1218(85) & 8.90 & side lobe & $\mathit{f_{70}}$ & 4.614637(27) & 0.1203(85) & 8.75 & side lobe \\ 
$\mathit{f_{71}}$ & 0.015675(52) & 0.120(17) & 4.47 & 7$\mathit{f_{7}}$$-$$\mathit{f_{31}}$ & $\mathit{f_{72}}$ & 5.062966(27) & 0.1193(85) & 8.66 & $\mathit{f_{13}}$$-$3$\mathit{f_{63}}$ \\ 
$\mathit{f_{73}}$ & 0.064300(52) & 0.118(16) & 4.51 & $\mathit{f_{7}}$$-$$\mathit{f_{51}}$ & $\mathit{f_{74}}$ & 7.457145(21) & 0.1179(66) & 10.98 & 4$\mathit{f_{5}}$$+$$\mathit{f_{65}}$ \\ 
$\mathit{f_{75}}$ & 0.287801(48) & 0.116(15) & 4.86 & side lobe & $\mathit{f_{76}}$ & 5.012063(28) & 0.1161(85) & 8.42 & 2$\mathit{f_{28}}$$-$$\mathit{f_{30}}$ \\ 
$\mathit{f_{77}}$ & 0.455877(47) & 0.114(14) & 4.94 & side lobe & $\mathit{f_{78}}$ & 1.543630(39) & 0.112(11) & 6.06 & 2$\mathit{f_{6}}$$-$$\mathit{f_{31}}$ \\ 
$\mathit{f_{79}}$ & 7.820432(23) & 0.1107(67) & 10.28 & 5$\mathit{f_{11}}$$+$$\mathit{f_{36}}$ & $\mathit{f_{80}}$ & 5.913888(25) & 0.1100(74) & 9.25 & $\mathit{f_{12}}$$-$$\mathit{f_{31}}$ \\ 
$\mathit{f_{81}}$ & 16.311333(30) & 0.1057(84) & 7.80 & side lobe & $\mathit{f_{82}}$ & 4.657617(30) & 0.1054(85) & 7.69 & 6$\mathit{f_{6}}$$-$$\mathit{f_{18}}$ \\ 
$\mathit{f_{83}}$ & 5.157936(31) & 0.1053(85) & 7.63 & 3$\mathit{f_{5}}$$+$2$\mathit{f_{44}}$ & $\mathit{f_{85}}$ & 4.971226(31) & 0.1037(85) & 7.52 & side lobe \\ 
$\mathit{f_{86}}$ & 3.184169(20) & 0.1037(54) & 11.83 & side lobe & $\mathit{f_{88}}$ & 5.027535(32) & 0.1015(85) & 7.36 & side lobe \\ 
$\mathit{f_{89}}$ & 7.431302(25) & 0.0988(66) & 9.20 & 8$\mathit{f_{11}}$$-$$\mathit{f_{23}}$ & $\mathit{f_{90}}$ & 4.654931(32) & 0.0985(85) & 7.19 & 6$\mathit{f_{6}}$$-$$\mathit{f_{27}}$ \\ 
$\mathit{f_{91}}$ & 5.155692(33) & 0.0980(85) & 7.10 & side lobe & $\mathit{f_{92}}$ & 1.513742(45) & 0.098(12) & 5.21 & 7$\mathit{f_{7}}$$-$$\mathit{f_{67}}$ \\ 
$\mathit{f_{93}}$ & 5.114072(33) & 0.0972(85) & 7.05 & side lobe & $\mathit{f_{94}}$ & 5.111386(34) & 0.0955(85) & 6.92 & $\mathit{f_{9}}$$+$$\mathit{f_{90}}$ \\ 
$\mathit{f_{95}}$ & 0.467336(57) & 0.095(14) & 4.12 & $\mathit{f_{9}}$$+$2$\mathit{f_{36}}$ & $\mathit{f_{96}}$ & 5.058205(34) & 0.0936(85) & 6.79 & 2$\mathit{f_{7}}$$+$$\mathit{f_{18}}$ \\ 
$\mathit{f_{97}}$ & 13.4290421(98) & 0.0924(24) & 23.92 & 6$\mathit{f_{15}}$$+$2$\mathit{f_{82}}$ & $\mathit{f_{98}}$ & 2.979607(24) & 0.0909(57) & 9.84 & 2$\mathit{f_{5}}$$-$$\mathit{f_{16}}$ \\ 
$\mathit{f_{99}}$ & 1.273069(56) & 0.090(13) & 4.20 & 6$\mathit{f_{7}}$$-$$\mathit{f_{54}}$ & $\mathit{f_{100}}$ & 4.569413(36) & 0.0897(85) & 6.55 & side lobe \\ 
$\mathit{f_{101}}$ & 1.752612(46) & 0.087(11) & 5.07 & 2$\mathit{f_{49}}$$+$$\mathit{f_{37}}$ & $\mathit{f_{102}}$ & 5.068202(38) & 0.0851(85) & 6.18 & $\mathit{f_{9}}$$+$$\mathit{f_{18}}$ \\ 
$\mathit{f_{103}}$ & 5.073507(39) & 0.0829(85) & 6.02 & 9$\mathit{f_{49}}$$-$$\mathit{f_{54}}$ & $\mathit{f_{104}}$ & 4.924948(39) & 0.0824(85) & 6.02 & side lobe \\ 
$\mathit{f_{105}}$ & 3.224768(25) & 0.0820(54) & 9.48 & side lobe & $\mathit{f_{106}}$ & 16.310279(39) & 0.0807(84) & 5.95 & side lobe \\ 
$\mathit{f_{107}}$ & 1.485247(55) & 0.080(12) & 4.23 & $\mathit{f_{15}}+7f_{\rm orb}$ & $\mathit{f_{108}}$ & 7.669935(33) & 0.0776(67) & 7.17 & 8$\mathit{f_{32}}$$+$$\mathit{f_{99}}$ \\ 
$\mathit{f_{109}}$ & 5.157188(43) & 0.0757(85) & 5.49 & side lobe & $\mathit{f_{110}}$ & 2.317604(50) & 0.0751(99) & 4.71 & 5$\mathit{f_{9}}$$+$$\mathit{f_{64}}$ \\ 
$\mathit{f_{111}}$ & 8.096399(32) & 0.0750(64) & 7.27 & $\mathit{f_{12}}$$+$4$\mathit{f_{51}}$ & $\mathit{f_{112}}$ & 4.972688(43) & 0.0745(85) & 5.40 & side lobe \\ 
$\mathit{f_{113}}$ & 4.958475(44) & 0.0739(85) & 5.36 & 6$\mathit{f_{5}}$$-$$\mathit{f_{52}}$ & $\mathit{f_{114}}$ & 2.125964(51) & 0.0738(100) & 4.58 & $\mathit{f_{3}}$$+$$\mathit{f_{7}}$ \\ 
$\mathit{f_{115}}$ & 4.700767(44) & 0.0728(85) & 5.33 & 3$\mathit{f_{6}}$$+$$\mathit{f_{54}}$ & $\mathit{f_{116}}$ & 5.019000(44) & 0.0724(85) & 5.26 & side lobe \\ 
$\mathit{f_{117}}$ & 3.268462(28) & 0.0715(54) & 8.26 & 3$\mathit{f_{5}}$$-$$\mathit{f_{24}}$ & $\mathit{f_{118}}$ & 3.182367(29) & 0.0714(54) & 8.15 & 8$\mathit{f_{9}}$$-$$\mathit{f_{95}}$ \\ 
$\mathit{f_{119}}$ & 4.967995(46) & 0.0702(85) & 5.09 & side lobe & $\mathit{f_{120}}$ & 7.818358(36) & 0.0700(67) & 6.50 & 3$\mathit{f_{6}}$$+$$\mathit{f_{60}}$ \\ 
$\mathit{f_{121}}$ & 14.765935(15) & 0.0695(28) & 15.26 & $\mathit{f_{1}}$$-$$\mathit{f_{6}}$ & $\mathit{f_{122}}$ & 9.040388(24) & 0.0695(44) & 9.74 & 2$\mathit{f_{11}}$$+$$\mathit{f_{80}}$ \\ 
$\mathit{f_{123}}$ & 2.979131(31) & 0.0694(57) & 7.51 & side lobe & $\mathit{f_{124}}$ & 9.365388(22) & 0.0688(40) & 10.55 & $\mathit{f_{6}}$$+$$\mathit{f_{79}}$ \\ 
$\mathit{f_{125}}$ & 7.732704(37) & 0.0683(67) & 6.35 & 9$\mathit{f_{15}}$$+$$\mathit{f_{11}}$ & $\mathit{f_{126}}$ & 5.012641(48) & 0.0672(85) & 4.88 & side lobe \\ 
$\mathit{f_{127}}$ & 7.262682(38) & 0.0670(68) & 6.10 & 9$\mathit{f_{32}}$$+$$\mathit{f_{73}}$ & $\mathit{f_{128}}$ & 4.356079(48) & 0.0661(83) & 4.90 & 5$\mathit{f_{3}}$$-$$\mathit{f_{83}}$ \\ 
$\mathit{f_{129}}$ & 11.720940(16) & 0.0646(27) & 15.04 & 4$\mathit{f_{5}}$$+$$\mathit{f_{23}}$ & $\mathit{f_{130}}$ & 2.738969(43) & 0.0639(73) & 5.43 & 5$\mathit{f_{9}}$$+$$\mathit{f_{35}}$ \\ 
$\mathit{f_{131}}$ & 5.001556(51) & 0.0635(85) & 4.60 & 6$\mathit{f_{5}}$$-$$\mathit{f_{57}}$ & $\mathit{f_{132}}$ & 8.097521(38) & 0.0633(64) & 6.14 & 2$\mathit{f_{11}}$$+$$\mathit{f_{57}}$ \\ 
$\mathit{f_{133}}$ & 7.480675(40) & 0.0632(67) & 5.83 & 7$\mathit{f_{16}}$$+$$\mathit{f_{103}}$ & $\mathit{f_{134}}$ & 5.014375(52) & 0.0625(85) & 4.53 & side lobe \\ 
$\mathit{f_{135}}$ & 5.105979(52) & 0.0622(85) & 4.51 & 8$\mathit{f_{3}}$$-$2$\mathit{f_{96}}$ & $\mathit{f_{136}}$ & 4.615623(52) & 0.0617(85) & 4.49 & $\mathit{f_{113}}-3f_{\rm orb}$ \\ 
$\mathit{f_{137}}$ & 17.855813(14) & 0.0594(23) & 16.29 & $\mathit{f_{1}}$$+$$\mathit{f_{6}}$ & $\mathit{f_{138}}$ & 3.184747(35) & 0.0582(54) & 6.64 & 2$\mathit{f_{7}}$$+$$\mathit{f_{130}}$ \\ 
$\mathit{f_{139}}$ & 3.225652(35) & 0.0572(54) & 6.61 & side lobe & $\mathit{f_{140}}$ & 6.928637(45) & 0.0569(67) & 5.23 & $\mathit{f_{3}}$$+$$\mathit{f_{87}}$ \\ 
$\mathit{f_{141}}$ & 5.062082(57) & 0.0566(85) & 4.11 & side lobe & $\mathit{f_{142}}$ & 4.862552(56) & 0.0565(84) & 4.16 & 6$\mathit{f_{9}}$$+$$\mathit{f_{114}}$ \\ 
$\mathit{f_{143}}$ & 4.882886(57) & 0.0560(84) & 4.13 & $\mathit{f_{94}}-2f_{\rm orb}$ & $\mathit{f_{144}}$ & 5.113392(58) & 0.0559(85) & 4.05 & side lobe \\ 
$\mathit{f_{145}}$ & 9.394903(27) & 0.0558(40) & 8.59 & 9$\mathit{f_{15}}$$+$$\mathit{f_{58}}$ & $\mathit{f_{146}}$ & 6.635804(43) & 0.0554(63) & 5.41 & 4$\mathit{f_{6}}$$+$$\mathit{f_{9}}$ \\ 
$\mathit{f_{147}}$ & 4.854698(57) & 0.0552(84) & 4.06 & $\mathit{f_{28}}-f_{\rm orb}$ & $\mathit{f_{148}}$ & 4.241761(57) & 0.0546(83) & 4.07 & $\mathit{f_{128}}-f_{\rm orb}$ \\ 
$\mathit{f_{149}}$ & 7.850966(46) & 0.0543(66) & 5.06 & 5$\mathit{f_{5}}$$-$$\mathit{f_{56}}$ & $\mathit{f_{150}}$ & 3.681427(43) & 0.0541(62) & 5.43 & $\mathit{f_{9}}$$+$$\mathit{f_{58}}$ \\ 
$\mathit{f_{151}}$ & 11.500193(19) & 0.0531(27) & 12.09 & 7$\mathit{f_{6}}$$+$$\mathit{f_{15}}$ & $\mathit{f_{152}}$ & 7.457621(48) & 0.0524(66) & 4.88 & side lobe \\ 
$\mathit{f_{153}}$ & 2.980049(41) & 0.0523(57) & 5.66 & side lobe & $\mathit{f_{154}}$ & 2.738459(53) & 0.0514(73) & 4.37 & side lobe \\ 
$\mathit{f_{155}}$ & 7.482953(50) & 0.0512(67) & 4.72 & 8$\mathit{f_{11}}$$-$$\mathit{f_{66}}$ & $\mathit{f_{156}}$ & 7.428582(50) & 0.0502(66) & 4.68 & 3$\mathit{f_{11}}$$+$$\mathit{f_{130}}$ \\ 
$\mathit{f_{157}}$ & 3.268972(40) & 0.0501(54) & 5.79 & side lobe & $\mathit{f_{158}}$ & 6.946965(51) & 0.0495(67) & 4.55 & 4$\mathit{f_{3}}$$-$$\mathit{f_{42}}$ \\ 
$\mathit{f_{159}}$ & 9.063374(34) & 0.0495(44) & 6.97 & 2$\mathit{f_{3}}$$+$3$\mathit{f_{101}}$ & $\mathit{f_{160}}$ & 9.288984(31) & 0.0495(41) & 7.45 & 3$\mathit{f_{6}}$$+$$\mathit{f_{90}}$ \\ 
$\mathit{f_{161}}$ & 3.820601(49) & 0.0492(63) & 4.79 & 2$\mathit{f_{3}}$$+$$\mathit{f_{71}}$ & $\mathit{f_{162}}$ & 3.821349(49) & 0.0490(63) & 4.77 & side lobe \\ 
$\mathit{f_{163}}$ & 7.819820(52) & 0.0481(67) & 4.47 & side lobe & $\mathit{f_{164}}$ & 6.305364(45) & 0.0473(56) & 5.19 & 6$\mathit{f_{3}}$$-$$\mathit{f_{94}}$ \\ 
$\mathit{f_{165}}$ & 14.403056(21) & 0.0471(26) & 11.27 & 6$\mathit{f_{11}}$$+$$\mathit{f_{87}}$ & $\mathit{f_{166}}$ & 9.396093(32) & 0.0469(40) & 7.22 & 7$\mathit{f_{11}}$$-$$\mathit{f_{78}}$ \\ 
$\mathit{f_{167}}$ & 7.319943(56) & 0.0453(67) & 4.16 & 8$\mathit{f_{16}}$$+$$\mathit{f_{84}}$ & $\mathit{f_{168}}$ & 3.183013(46) & 0.0444(54) & 5.07 & side lobe \\ 
$\mathit{f_{169}}$ & 6.926223(58) & 0.0439(67) & 4.04 & 4$\mathit{f_{3}}$$-$$\mathit{f_{15}}$ & $\mathit{f_{170}}$ & 3.793637(54) & 0.0438(63) & 4.29 & 6$\mathit{f_{5}}$$-$2$\mathit{f_{47}}$ \\ 
$\mathit{f_{171}}$ & 7.340991(58) & 0.0437(67) & 4.04 & 8$\mathit{f_{6}}$$-$$\mathit{f_{66}}$ & $\mathit{f_{172}}$ & 3.640488(50) & 0.0436(58) & 4.63 & $\mathit{f_{9}}$$+$$\mathit{f_{60}}$ \\ 
$\mathit{f_{173}}$ & 9.397589(35) & 0.0435(40) & 6.69 & 7$\mathit{f_{11}}$$-$$\mathit{f_{78}}$ & $\mathit{f_{174}}$ & 3.620868(51) & 0.0427(58) & 4.57 & $\mathit{f_{3}}$$+$$\mathit{f_{24}}$ \\ 
$\mathit{f_{175}}$ & 8.091639(57) & 0.0419(64) & 4.06 & 2$\mathit{f_{6}}$$+$$\mathit{f_{131}}$ & $\mathit{f_{176}}$ & 3.639944(53) & 0.0418(58) & 4.44 & side lobe \\ 
$\mathit{f_{177}}$ & 3.822063(58) & 0.0414(63) & 4.04 & 2$\mathit{f_{3}}$$+$$\mathit{f_{71}}$ & $\mathit{f_{178}}$ & 6.328588(53) & 0.0403(57) & 4.41 & 5$\mathit{f_{3}}$$-$$\mathit{f_{138}}$ \\ 
$\mathit{f_{179}}$ & 3.183557(51) & 0.0399(54) & 4.56 & side lobe & $\mathit{f_{180}}$ & 6.331410(54) & 0.0398(57) & 4.36 & 5$\mathit{f_{3}}$$-$$\mathit{f_{60}}$ \\ 
$\mathit{f_{181}}$ & 6.307914(54) & 0.0395(56) & 4.32 & 6$\mathit{f_{7}}$$+$$\mathit{f_{28}}$ & $\mathit{f_{182}}$ & 3.224258(52) & 0.0388(54) & 4.48 & side lobe \\ 
$\mathit{f_{183}}$ & 18.213762(20) & 0.0377(20) & 11.58 & $\mathit{f_{1}}$$+$$\mathit{f_{3}}$ & $\mathit{f_{184}}$ & 9.006555(45) & 0.0374(44) & 5.21 & $\mathit{f_{12}}$$+$$\mathit{f_{31}}$ \\ 
$\mathit{f_{185}}$ & 8.492431(56) & 0.0372(55) & 4.18 & $\mathit{f_{1}}$$-$$\mathit{f_{120}}$ & $\mathit{f_{186}}$ & 3.267986(56) & 0.0363(54) & 4.20 & side lobe \\ 
$\mathit{f_{187}}$ & 11.723354(28) & 0.0362(27) & 8.43 & $\mathit{f_{7}}$$+$$\mathit{f_{151}}$ & $\mathit{f_{188}}$ & 3.226706(56) & 0.0362(54) & 4.18 & 3$\mathit{f_{49}}$$+$$\mathit{f_{92}}$ \\ 
$\mathit{f_{189}}$ & 18.257354(21) & 0.0360(20) & 11.12 & 8$\mathit{f_{3}}$$+$2$\mathit{f_{38}}$ & $\mathit{f_{190}}$ & 8.530583(58) & 0.0349(54) & 4.01 & 5$\mathit{f_{28}}$$-$$\mathit{f_{8}}$ \\ 
$\mathit{f_{191}}$ & 3.447827(58) & 0.0345(53) & 4.02 & $\mathit{f_{3}}$$+$$\mathit{f_{6}}$ & $\mathit{f_{192}}$ & 9.367870(45) & 0.0339(40) & 5.20 & 6$\mathit{f_{32}}$$+$$\mathit{f_{84}}$ \\ 
$\mathit{f_{193}}$ & 11.497779(31) & 0.0334(27) & 7.60 & 8$\mathit{f_{49}}$$+$$\mathit{f_{140}}$ & $\mathit{f_{194}}$ & 9.289630(47) & 0.0329(41) & 4.96 & side lobe \\ 
$\mathit{f_{195}}$ & 8.805122(58) & 0.0320(49) & 4.05 & 8$\mathit{f_{9}}$$+$$\mathit{f_{33}}$ & $\mathit{f_{196}}$ & 8.926036(55) & 0.0311(45) & 4.24 & $\mathit{f_{122}}-f_{\rm orb}$ \\ 
$\mathit{f_{197}}$ & 9.366340(49) & 0.0310(40) & 4.75 & side lobe & $\mathit{f_{198}}$ & 9.393202(51) & 0.0299(40) & 4.60 & 9$\mathit{f_{3}}$$-$$\mathit{f_{125}}$ \\ 
$\mathit{f_{199}}$ & 9.363416(51) & 0.0296(40) & 4.53 & $\mathit{f_{1}}$$-$$\mathit{f_{158}}$ & $\mathit{f_{200}}$ & 14.643762(34) & 0.0295(27) & 6.81 & 6$\mathit{f_{3}}$$+$$\mathit{f_{188}}$ \\ 
$\mathit{f_{201}}$ & 9.028079(58) & 0.0289(44) & 4.04 & 9$\mathit{f_{3}}$$-$$\mathit{f_{132}}$ & $\mathit{f_{202}}$ & 11.721552(36) & 0.0277(27) & 6.44 & side lobe \\ 
$\mathit{f_{203}}$ & 11.720090(36) & 0.0277(27) & 6.44 & side lobe & $\mathit{f_{204}}$ & 9.417073(55) & 0.0276(40) & 4.27 & 2$\mathit{f_{32}}$$+$$\mathit{f_{120}}$ \\ 
$\mathit{f_{205}}$ & 9.394325(55) & 0.0274(40) & 4.21 & side lobe & $\mathit{f_{206}}$ & 11.106405(40) & 0.0269(28) & 5.87 & 4$\mathit{f_{6}}$$+$$\mathit{f_{30}}$ \\ 
$\mathit{f_{207}}$ & 17.972987(31) & 0.0260(21) & 7.48 & $\mathit{f_{1}}$$+$$\mathit{f_{5}}$ & $\mathit{f_{208}}$ & 9.727995(54) & 0.0255(36) & 4.35 & 6$\mathit{f_{6}}$$+$$\mathit{f_{35}}$ \\ 
$\mathit{f_{209}}$ & 14.620130(43) & 0.0232(27) & 5.39 & 5$\mathit{f_{3}}$$+$$\mathit{f_{135}}$ \\  
\end{longtable}


\end{document}